\documentstyle[prc,preprint,aps]{revtex}
\begin{document}
\draft
\title{Isospin-mixing corrections for {\it fp}-shell Fermi transitions}
\author{W.~E. Ormand}
\address{Physics Department, 401 Nielsen Hall, University of Tennessee,\\
Knoxville, TN 37996-1200\\
 and \\
Physics Division, Oak Ridge National Laboratory, P.O. Box 2008, \\
MS-6373 Building 6003, Oak Ridge, TN 37831-6373}
\author{B.~A. Brown}
\address{National Superconducting Cyclotron Laboratory and
Department of Physics and Astronomy\\
Michigan State University, East Lansing, MI 48824-1321}
\maketitle
\begin{abstract}
Isospin-mixing corrections for superallowed Fermi transitions
in {\it fp}-shell nuclei are computed within the framework of the shell model.
The study includes three nuclei that are part of the set of nine accurately
measured transitions as well as five cases that are expected to be measured in
the future at radioactive-beam facilities. We also include some new
calculations
for $^{10}$C. With the isospin-mixing corrections applied to the nine
accurately measured $ft$ values, the conserved-vector-current hypothesis
and the unitarity condition of the Cabbibo-Kobayashi-Maskawa (CKM) matrix are
tested.
\end{abstract}
\pacs{PACS numbers.23.40.-s }

Superallowed Fermi $\beta$ transitions in nuclei,
$(J^\pi=0^+,T=1)\rightarrow (J^\pi=0^+,T=1)$,
provide an excellent laboratory for
precise tests of the properties of the electroweak interaction, and have been
the subject of intense study for several decades (cf.
Refs.~\cite{ref1,Tow77,Orm89,Har90,Sir78}). According to
the conserved-vector-current (CVC) hypothesis, the $ft$ values for pure Fermi
transitions should be nucleus independent, and given by
\begin{equation}
ft=\frac{K}{G_V^2|M_F|^2},
\end{equation}
where $K/(\hbar c)^6=2\pi^3\ln 2 \hbar /(m_ec^2)^5=
8.120270(12)\times 10^{-7}~{\rm GeV}^{-4}{\rm s}$, $G_V$
is the vector coupling constant for nuclear $\beta$
decay, and $M_F$ is the Fermi
matrix element, $M_F=\langle\psi_f\mid T_\pm \mid\psi_i\rangle$. By
comparing the decay rates for muon and nuclear Fermi $\beta$ decay, the
Cabbibo-Kobayashi-Maskawa (CKM) mixing matrix element~\cite{CKM} between $u$
and $d$
quarks ($v_{ud}$) can be determined
and a precise test of the unitarity
condition of the CKM matrix under the assumption of the three-generation
standard model is possible~\cite{Sir78,CKM}.

For tests of the standard model, two
nucleus-dependent corrections must be applied to experimental $ft$
values. The first is a series of radiative corrections to the statistical
rate function $f$, embodied in the factors $\delta_R$ and $\Delta_R$, giving
$f_R=f(1+\delta_R+\Delta_R)$~\cite{Bli73,Sir86,Jau90,Bar92,Tow92,foota}.
Where $\delta_R$
is due to standard, electromagnetic (``inner'')
radiative corrections (cf. p.~45 in Ref.~\cite{Bli73}), while
$\Delta_R$ is what has been referred to as the ``outer'' radiative
correction (cf. p. 47 of Ref.~\cite{Bli73}) and includes axial-vector
interference
terms~\cite{Jau90,Bar92,Tow92}. The second correction is applied to the Fermi
matrix
element $M_F$, and is due to the presence of isospin-nonconserving (INC)
forces in nuclei, and is denoted by $\delta_C$~\cite{Tow77,Orm89,Tow89};
namely $\mid M_F\mid^2=\mid M_{F0}\mid^2(1-\delta_C)$, where
$M_{F0}=[T(T+1)-T_{Z_i}T_{Z_f}]^{1/2}$.

With the ``nucleus-independent''
${\cal F}t$ values defined by
\begin{equation}
{\cal F}t=ft(1+\delta_R+\Delta_R)(1-\delta_C),
\end{equation}
the CKM matrix element $v_{ud}$ is given by~\cite{Bar92}
\begin{equation}
\mid v_{ud}\mid^2 = \frac{\pi^3\ln 2}{{\cal F}t}\frac{\hbar^7}{G_F^2m_e^5c^4}
=\frac{2984.38(6)~{\rm s}}{{\cal F}t},
\end{equation}
where the Fermi coupling constant, $G_F$ is obtained from muon $\beta$-decay,
and includes radiative corrections.
Currently, $ft$ values for nine superallowed transitions have been measured
with an experimental precision of better than 0.2\%~\cite{Har90,Sav95}.
Prior to the recent measurement for $^{10}$C, the experimental $ft$-values
gave some hint of an additional $Z$ dependence not presently accounted for.
In addition, the unitarity condition for the CKM matrix was not satisfied.
This prompted studies to empirically determine the ``missing''
correction and to satisfy the CVC requirement~\cite{Wil93}.
Recent results for $^{10}$C~\cite{Sav95}, however, do not support the
conclusion that there may be a ``missing'' correction, as together all nine
${\cal F}t$ values satisfy the constancy requirement of the CVC hypothesis. The
unitarity condition of the CKM matrix, however, is still violated at the
level of $\sim~3\sigma$~\cite{Bar92,Tow92,Sav95}, and can
only be restored by the
application of an across the board correction of approximately 0.3-0.4\%.
In the future, a possible $Z$ dependence in the ${\cal F}t$ values can be
further tested by a remeasurement of $^{10}$C and precise measurements of
heavier
{\it fp}-shell Fermi transitions using radioactive beams.

The necessary formalism for computing $\delta_C$ is given in
Refs.~\cite{Tow77,Orm85}, and conventionally, $\delta_C$ is factored into
two components, i.e., $\delta_C=\delta_{IM}+\delta_{RO}$~\cite{Tow77}.
The correction $\delta_{IM}$ is due to isospin mixing between different
valence shell-model configuration states (eg., the $0\hbar\omega$ $1s0d$
shell). The essential ingredients for $\delta_{IM}$
are a base isoscalar shell-model Hamiltonian that reproduces the spectra of
excited $J=0$ states and an INC interaction that reproduces experimental mass
splittings~\cite{Orm85}.
The second correction, $\delta_{RO}$, is due to the
deviation from unity of the radial overlap between the converted proton and
the corresponding neutron. This effect corresponds to the influence of states
that lie outside the valence shell-model configuration space
(eg., $2\hbar\omega$, one particle-one hole configurations).
Currently, there are two approaches for
evaluating $\delta_{RO}$ that give roughly the same agreement
with the CVC hypothesis, but are in overall disagreement in magnitude.
In the first approach~\cite{Tow77}, the radial wave
functions were obtained using a Woods-Saxon (WS) plus Coulomb potential,
while in the second~\cite{Orm89,Orm85}, self-consistent Hartree-Fock (HF)
calculations using Skyrme-type interactions (including Coulomb) were performed.
The principal feature of the HF procedure is that since the mean field is
proportional
to the nuclear densities, the Coulomb force induces a one-body isovector
potential that tends to counter Coulomb repulsion, therefore reducing
$\delta_{RO}$. Because of this, the HF values of $\delta_{RO}$ are
consistently smaller than the WS values by approximately 0.1-0.2 (in \%).

In this paper, we re-evaluate the isospin-mixing corrections for the
{\it fp}-shell transitions $^{46}$V, $^{50}$Mn, and $^{54}$Co that are included
in
the set of nine accurately measured transitions using expanded shell-model
spaces and improved effective interactions. Comparisons with experimental data
on the isospin-forbidden transition to the first excited ($J=0,T=1)$
state, which places some constraints on $\delta_{IM}$~\cite{Hag94}, will
also be made.
In addition, one application of future radioactive beam facilities is to
extend the data set to the heavier {\it fp}-shell nuclei $^{58}$Zn,
$^{62}$Ga, $^{66}$As, $^{70}$Br,
and $^{74}$Rb~\cite{foot0}. Such a study may shed light on any possible $Z$
dependence in
the ${\cal F}t$ values. As such, we present calculations for the important
isospin-mixing corrections for these nuclei. We find for these nuclei that
both $\delta_{IM}$ and $\delta_{RO}$ are much larger than in the case of the
previous nine transitions. In addition, the
difference between the Woods-Saxon and Hartree-Fock calculations for
$\delta_{RO}$ is more pronounced for these nuclei, and
precise measurements of these cases may be able to make a selection
between the two approaches.

A calculation of $\delta_C$ begins with defining the shell-model configuration
space and the base isoscalar shell-model Hamiltonian. Naturally, these are
not independent choices, as model-space truncations may require
renormalizations of the effective interaction.
For the nuclei under consideration here,
the base configuration space is comprised of the $0f_{7/2}$, $1p_{3/2}$,
$1p_{1/2}$, and $0f_{5/2}$ orbitals, or {\it fp} shell. Because of
computational restrictions, some model space truncations must be imposed on all
nuclei except $^{46}$V and $^{74}$Rb. The active model space used for each
nucleus
is listed in Table I. These model-space truncations were found to be adequate
except for the cases of $A=54$ and 74 as discussed below.
In recent years, progress has been
made towards the determination of effective interactions for use in
{\it fp}-shell calculations, in particular for the lower part of the
shell~\cite{Ric90}. In this work, the FPD6 interaction of Ref.~\cite{Ric90}
was used for $A \le 50$. For $A=54$ the interaction was taken to be
comprised of the two-body matrix elements of FPD6, while the single-particle
energies were renormalized to reproduce the experimental binding energies
of $^{57}$Ni assuming a closed $f_{7/2}$ core (FPD6$^*$).
In the upper part of the {\it fp} shell, the interaction is less well
determined, and for $58\le A \le 74$, we compare the results obtained
using FPD6$^*$ and the FPVH interaction of Ref.~\cite{FPVH}.
The calculations presented here were performed using a
unix version of the shell-model code OXBASH~\cite{oxbash}
on Silicon Graphics computers at Oak Ridge National Laboratory.

Another popular interaction used recently, but not here
for the reasons outlined below, is a
modified version of the original Kuo-Brown interaction referred to as
KB3~\cite{Pov81}. Although this interaction gives very nearly the same
results as FPD6 and FPD6$^*$ in the lower {\it fp} shell, it begins to diverge
drastically from either FPD6$^*$ or FPVH for $A\ge 60$. The reason for this is
that in the upper part of the shell, monopole terms in KB3 tend to push
the $0f_{5/2}$ orbit up, creating a large gap between the {\it p} orbitals
and the $0f_{5/2}$ orbit. In fact, for the single-hole nucleus $A=79$, KB3
predicts the ground state to be $J^\pi=5/2^-$ with excitation energies
for the $1/2^-$ and $3/2^-$ hole states of 3.753~MeV and 7.010~MeV,
respectively. This is in strong disagreement with spherical
Hartree-Fock calculations, where, for example, the Skyrme M$^*$
force~\cite{Bar82}
predicts the ground
state to be $J^\pi=1/2^-$,
with excitation energies for the $5/2^-$ and $3/2^-$ hole states to be
0.591~MeV and 1.460~MeV, respectively. Both FPD6$^*$ and FPVH are in excellent
agreement with the HF results.

To evaluate the configuration-mixing contribution $\delta_{IM}$
we use an INC interaction derived in the same manner as in Ref.~\cite{Orm89a}.
An important ingredient of the INC interaction is the mass scaling of the
Coulomb
two-body strength and single-particle energies as governed by the oscillator
parameter $\hbar\omega$ (cf. Eq.~(3.5) of Ref.~\cite{Orm89a}). Since there are
important deviations from the usual smooth formulae for $\hbar\omega$ around
$A\sim
53-59$, and we want a uniform parameterization across the {\it fp} shell, we
have
chosen $\hbar\omega$ so as to reproduce the rms point proton radii obtained
from
with a spherical Hartree-Fock calculation using the Skyrme M$^*$
force.
The values
of $\hbar\omega$ used here are listed in Table I. Using these values of
$\hbar\omega$, the parameters of the INC interaction of Ref.~\cite{Orm89a}
were redetermined. In addition, the single-particle energies of the $0f_{5/2}$
and
$1p_{1/2}$ orbits were not well determined by the data set in
Ref.~\cite{Orm89a},
and were chosen to reproduce the Coulomb splittings
for the $J^\pi=5/2^-$ and $1/2^-$ $A=57$, $T=1/2$
multiplets~\cite{Bha92} assuming a closed $^{56}$Ni core. The
parameters of the INC interaction used are $\epsilon(0f_{7/2})=7.487$~MeV,
$\epsilon(1p_{3/2})=7.312$~MeV,
$\epsilon(0f_{5/2})=7.582$~MeV, $\epsilon(1p_{1/2})=7.240$~MeV, $S_C=1.006$,
$S_0^{(1)}=0.0$, and $S_0^{(2)}=-4.2\times 10^{-2}$.

Shown in Tables II (FPD6$^*$ for $A\ge 58$) and III (FPVH for $A\ge 58$) are
the
results of shell-model calculations for
$\delta_{IM}$ for the {\it fp}-shell nuclei under consideration.
In addition,
the theoretical and experimental values for the excitation energy of the
first excited $J^\pi=0^+,T=1$ state are shown. Generally, for $A<58$
one finds that $\delta_{IM}$ is of the order 0.02-0.10\%,
while for the heavier nuclei it
can be as large as 0.4\%.
One reason for the increase in $\delta_{IM}$ for $A \ge 62$ is that the
excitation energy of the lowest $J=0, T=0$ state is steadily decreasing in
these nuclei, eventually becoming equal to or less than that for the $J=0,T=1$
state. The effect of $T=0$ mixing in the $T_z=0$ parent is to remove Fermi
strength from the transition, therefore increasing $\delta_{IM}$.
The second reason for the enhancement in $\delta_{IM}$ is that the excitation
energy of the first excited $J=0,T=1$ state is lower in these nuclei than
for $A\le 54$. The contribution to
$\delta_{IM}$ due to mixing with this state is given by
\begin{equation}
\delta_{IM}^1=[\alpha(0)-\alpha(-1)]^2,
\label{dim}
\end{equation}
where $\alpha(T_z)$ is the amplitude for mixing the first excited state into
the ground state for the nucleus with third component of isospin
$T_z=(Z-N)/2$, ($Z$ and $N$ denoting the number of protons and neutrons,
respectively). In perturbation theory, the mixing amplitude $\alpha$ is
is determined by the ratio of the matrix element of the INC interaction and
the energy difference between the states, i.e.
\begin{equation}
\alpha = \langle \psi_1 | V_{INC} | \psi_0 \rangle/\Delta E_{01}.
\end{equation}
Therefore, a dependence in $\delta_{IM}$ on the isoscalar interaction and
shell-model configuration space is manifested in the reproduction of the
energy spectrum of $J=0$ states. Improved values for $\delta_{IM}^1$ and
$\delta_{IM}$ maybe obtained by scaling $\delta_{IM}^1$ by the square of the
ratio of the theoretical and experimental excitation energies,
$(\Delta E_{01}^{th}/\Delta E_{01}^{exp})^2$. The results are tabulated in
Tables II and III with the additional subscript $s$.
In addition, for $^{46}$V the contribution due to the second excited state,
$\delta_{IM}^2=0.012\%$ was also scaled by the ratio $(5.84/3.57)^2$
to account for
the difference between the experimental and theoretical excitation energies
for this state as well. As is pointed out in Ref.~\cite{Hag94},
the experimentally measured Fermi matrix element for the
isospin-forbidden transition from the ground state of the parent to the first
excited $J=0,T=1$ state in the daughter can be related to
$\delta_{IM}^1$~\cite{footb}.
The experimental and theoretical values are compared in Table II, where overall
good agreement is achieved except for $A=54$.

Two nuclei in this study deserve special mention in regards to
model-space truncations. The first is $A=74$. Towards the
upper end of the {\it fp} shell, it is apparent that deformation effects are
beginning to become
important as can be seen by the steady decrease with nucleon number $A$ in the
excitation energy of the lowest $J^\pi=2^+$ states in even-even $N=Z$
nuclei~\cite{Led78,Lis88} as
shown in Table IV. Also shown in Table IV is a comparison between the
experimental excitation energies and those obtained from a shell-model
calculation using
the FPD6$^*$ and FPVH interactions. A clear change is observed between $A$=72
and 76, and for this reason, a
proper calculation for $A=74$ should probably include the
$0g_{9/2}$ orbit. At present such a calculation is not
feasible, and we express caution regarding the results for $A=74$ and the hope
that more thorough calculations can be performed in the near future.
The second case is $A=54$, where, to first order, the ground-state wave
function is comprised of two $f_{7/2}$ holes. Excited $J=0$ states, which
are important for $\delta_{IM}$, have at least two particles excited outside of
the
$0f_{7/2}$ orbit (i.e., a two particle-four hole ($2p-4h$) configuration
relative to
the $^{56}$Ni closed shell).
The effect of including these configurations, however, is to
decrease the binding energy of the ground state relative to the $2p-4h$ states,
leading to an artificially large excitation energy for the excited states. In
principle, if computational limitations permitted,
the inclusion of $4p-6h$ states would decrease this gap. A calculation
utilizing no restrictions with the $0f_{7/2}$ and $1p_{3/2}$ orbits is
feasible,
and the gap between the ground state and excited states is reduced
considerably.
The effects of isospin mixing in this space, however, are quite small, and are
in
disagreement the experimental results obtained in Ref.~\cite{Hag94}. In
addition,
when excitations involving two particles into the $0f_{5/2}$ orbit are
included,
the gap worsens, indicating that $4p-6h$ excitations to the $0f_{5/2}$ orbit
are important for describing the energy of the first excited state. An
alternative
approach is that of Ref.~\cite{Tow89} where the isoscalar interaction was
renormalized
in the $2p-4h$ space so that the excitation spectrum had the correct energies.
In
that work, $\delta_{IM}^1$ and $\delta_{IM}$ were
found to be 0.037(8)\% and 0.045(5)\%, respectively, and are in good agreement
with the experimental value for $\delta_{IM}^1$ of 0.035(5).
Given the computational limitations and
the experimental data, probably the best value of $\delta_{IM}$ for
$^{54}$Co when testing of CVC and the unitarity of the CKM matrix is 0.04(1)\%.

The radial overlap correction $\delta_{RO}$ was evaluated using the procedures
outlined in Refs.~\cite{Tow77,Orm85}. Shown in
Tables II (FPD6$^*$) and III (FPVH) are the results for
$\delta_{RO}$ using Hartree-Fock (HF) and Woods-Saxon (WS) single-particle
wave functions. The HF results were computed using the
Skyrme M$^*$ force~\cite{Bar82},
which generally gives better overall agreement with many experimental
observables than do other Skyrme forces, in particular some isovector
quantities
such as the centroid energies for giant-dipole and giant isovector-monopole
resonances~\cite{Gle90}. Therefore, we have chosen to present all the results
with
Skyrme M$^*$. However, we believe the dependence on the parameters of the
Skyrme
interaction should be further investigated~\cite{footc}. The WS values for
$A\ge 58$
were computed using the Woods-Saxon parameters given in Ref.~\cite{Str82}.

An interesting feature of $\delta_{RO}$ is that it is much larger for the
$A\ge 58$ cases. This is primarily due to:
(1) the larger difference between the
proton and neutron separation energies $\sim 10$~MeV; (2) the last proton being
rather weakly bound $\sim 2.5$~MeV, as opposed to 5-6~MeV for
$A \le 54$; and (3) $\delta_{RO}$ being dominated by
the $0p_{3/2}$ orbit, which has a lower centrifugal barrier than in the case
for
$A \le 54$, which is dominated by the $0f_{7/2}$ orbit. Finally, it is apparent
from Tables II and III that the difference between the HF and WS evaluations of
$\delta_{RO}$ is considerably larger for the heavier nuclei, ranging from
0.3-0.7\%, as opposed to 0.02-0.2\% for the
$A\le 54$ cases (cf., Ref.~\cite{Orm89}).
As such, CVC tests including accurate measurements of the
{\it ft} values for the heavier {\it fp}-shell cases may lead to a
differentiation between the two approaches.

To complete the survey of isospin-mixing corrections for Fermi transitions, the
values of $\delta_{IM}$ and $\delta_{RO}^{HF}$ (and the sum $\delta_C$)
for the nine accurately measured nuclei
are listed in Table V. The $\delta_{RO}^{HF}$ values were obtained
using the Skyrme M$^*$ force.
The values presented for $^{10}$C were evaluated using
the full $0p_{3/2},0p_{1/2}$ shell-model space and the CKPOT isoscalar
interaction~\cite{Coh65} and the INC interaction of Ref.~\cite{Orm89}.

Aside from the systematic difference between the HF and WS estimates of
$\delta_{RO}$ the theoretical uncertainty in $\delta_C$ for $A\le 54$ is of the
order 0.09\% in most cases~\cite{Orm89}. This arises from the addition in
quadrature of
0.05\% for $\delta_{IM}$, 0.06\% for $\delta_{RO}$, and 0.05\% as a
conservative
estimate for the spectator mismatch, which as discussed in
Refs.~\cite{Orm89,Wil95} is expected to be negligible. For $A\ge 58$ there are
some
differences between the results obtained with the FPD6$^*$ and FPVH
interactions.
For the most part, the $\delta_{IM}$ values are in overall agreement with
differences of the order 0.05\%. For $\delta_{RO}$ the mean difference
between the two interactions is 0.124\%, but can be as large 0.33\%. These
differences are primarily attributed to differences in the excitation energies
of
the $T=3/2$ states in the $A-1$ parent. For more precise studies in the future,
it will be necessary to improve upon the base shell-model isoscalar
interaction.
Nonetheless, both interactions predict large differences between the HF and WS
approaches to $\delta_{RO}$.

A test of the CVC hypothesis is performed by applying $\delta_C$ to the
$f_Rt$ values, which are also listed in Table V. Here, $f_Rt$
was computed by applying the radiative corrections listed in column 1 of Table
3 in
Ref.~\cite{Bar92} and the average of the $(\alpha/\pi)C_{NS}$ corrections
listed in in Refs.~\cite{Bar92,Tow92} to the $ft$ values of the new
Chalk River compilation~\cite{Tow95}.
Applying $\delta_C$ to $f_Rt$ (note that the ${\cal F}t$ are also listed in
Table
V) and taking the error-weighted average, we find
${\cal F}_{avg}t=3150.8\pm 1.2\pm 2.5$~s with $\chi^2/\nu=0.66$.
Using Eq.~(3) and $v_{us}=0.2199(17)$~\cite{Bar92}
and $v_{ub} < 0.0075$ (90\% confidence level)~\cite{Tho88}, the unitarity
condition of the CKM matrix is found to be
$0.9956\pm(0.0008)_{stat}\pm(0.0007)_{sys}$.
Thus, from the constancy of the ${\cal F}t$ values, we conclude that CVC
hypothesis
is satisfied, but that the unitarity condition of the CKM matrix is violated at
the level of 3-4 $\sigma$, and can only be
achieved with an additional negative correction of 0.3-0.4\% applied
uniformly to each nucleus. It is important to note that a correction of this
magnitude lies well outside the range of acceptable uncertainties in the
nuclear corrections.

In summary, the isospin-mixing corrections for Fermi transitions in
{\it fp}-shell nuclei were evaluated. The evaluation also included transitions
involving heavier nuclei that are expected to be measured in the future
radioactive-beam facilities. It was found that the isospin-mixing
corrections were considerably larger for the $A \ge 58$ cases. In addition, the
difference between the Hartree-Fock and Woods-Saxon method of evaluating
$\delta_{RO}$ was much larger for these nuclei. As such, accurate measurements
of
the $ft$-values for these nuclei might lead to a discrimination between the two
methods. In regard to the accurately measured transitions, it was found that
the
newer evaluations give better agreement with experiment for the
configuration-mixing term $\delta_{IM}$, with the noted exception of $^{54}$Co,
which poses a significant computational challenge. Lastly, it is found that the
corrected ${\cal F}t$ values are in excellent agreement with the CVC
hypothesis, but that
the unitarity condition of the CKM matrix is violated at the level of
3-4~$\sigma$.

\begin{center}
{\bf Acknowledgments}
\end{center}
We wish to thank I.~S.~Towner for providing us the Chalk River
$ft$ value data set, and for comments regarding this manuscript.
Oak Ridge National Laboratory is managed for the U.S. Department of Energy
by Martin Marietta Energy Systems, Inc. under contract No.
DE--AC05--84OR21400. Theoretical nuclear physics research at the University of
Tennessee is supported by the U.S. Department of Energy through contract No.
DE--FG05--93ER40770. BAB acknowledges support from the National Science
Foundation through grant no. PHY94-03666.

\newpage
\bibliographystyle{try}


\begin{thebibliography}{11}
\bibitem{foota} An entirely equivalent representation that keeps the
historical separation of the ``outer'' radiative corrections is given in
Ref.~\cite{Tow92}.
\bibitem{foot0} Although the ground-state angular momentum has not been
identified in some of the $N=Z$ cases, the short lifetimes of 100-200~ms
are consistent with a superallowed fermi transition.
\bibitem{footb} The leading order contribution to
the isospin-forbidden Fermi matrix element is Eq.(4). Higher-order terms
enter at the level of $\sim 0.002-0.005$\%. These two effects lead to a
near cancellation in the unscaled results for $^{54}$Co.
\bibitem{footc} The $\delta_{RO}$ given in Ref.~\cite{Orm89} are the
average of the values obtained
with Skyrme A, SGI, SGII, Skyrme M and M$^*$, and are generally
0.05-0.1\% smaller than the values in Tables II and III.
\bibitem{footd} The $ft$ value of $^{50}$Mn was modified to include
the two observed Gamow-Teller branches with strength 570 and
6.8 ppm reported in Ref.~\cite{Hag94}. In addtion, the $ft$
values listed in Ref.~\cite{Har90} are consistent with those in
Ref.~\cite{Bar92}.
\end{thebibliography}

\begin{table}
\caption{List of shell-model configuration spaces and $\hbar\omega$ used
for each nucleus}
\begin{tabular}{ccc}
Nucleus & Configuration & $\hbar\omega$ (MeV) \\
\tableline
$^{46}$V & full {\it fp} & 10.952\\
$^{50}$Mn & $(f_{7/2},p_{3/2})^{10}$
+ $f_{7/2}^{n_7},f_{5/2}^{n_5},p_{1/2}^{n_1}$ $(n_5+n_1=1)$ & 10.550 \\
$^{54}$Co & $(f_{7/2},p_{3/2})^{14}$
 + $f_{7/2}^{n_7},p_{3/2}^{n_3},f_{5/2}^{n_5},p_{1/2}^{n_1}$
$(n_3+n_5+n_1=2)$ & 10.486 \\
$^{58}$Zn & $f_{7/2}^{n_7},p_{3/2}^{n_3}
f_{5/2}^{n_5},p_{1/2}^{n_1}$ ($14 \le n_7 \le 16$) & 10.298\\
$^{62}$Ga & $f_{7/2}^{16},(p_{1/2},f_{5/2},p_{1/2})^6$ & 10.017\\
$^{66}$As & $f_{7/2}^{16},(p_{1/2},f_{5/2},p_{1/2})^{10}$ & 9.681 \\
$^{70}$Br & $f_{7/2}^{16},(p_{1/2},f_{5/2},p_{1/2})^{14}$ & 9.424 \\
$^{74}$Ga & full {\it fp} & 9.203 \\
\end{tabular}
\end{table}

\newpage
\begin{table}
\caption{List of isospin-mixing corrections $\delta_{IM}$ and
$\delta_{RO}$ (in \%), theoretical and experimental excitation energies for
the first $J=0,T=1$ excited state (in MeV), theoretical and
experimental values of $\delta_{IM}^1$. Values of $\delta_{IM}$ obtained by
setting the theoretical excitations equal to experiment are indicated by
the additional subscript $s$. Values of
$\delta_{RO}$ for Hartree-Fock and Woods-Saxon wave functions are denoted
by the superscripts HF and WS, respectively. The results obtained for
$A \ge 58$ are shown for the FPD6$^*$ interaction.}
\begin{tabular}{cccccccccc}
A & $E_{x,th}^1$ & $E_{x,exp}^1$ & $\delta_{IM}^1$ & $\delta_{IM,s}^1$ &
$\delta_{IM,exp}^1$ & $\delta_{IM}$ & $\delta_{IM,s}$ &
$\delta_{RO}^{{\rm HF}}$ &  $\delta_{RO}^{{\rm WS}}$\\
\tableline
$^{46}$V & 4.295 & 2.611 & 0.020 & 0.054 & 0.053(5)$^a$ & 0.040 & 0.094 &
0.286 & 0.36(6)$^b$ \\
$^{50}$Mn & 3.620 & 3.69 & 0.014 & 0.015 & $<$0.016$^a$ & 0.026 & 0.017 &
0.325 & 0.40(9)$^b$\\
$^{54}$Co & 6.423 & 2.561 & 0.0004 & 0.003 & 0.035(5)$^a$ & 0.003 & 0.006 &
0.397 & 0.56(6)$^b$\\
$^{58}$Zn & 2.850 & 2.943 & 0.196 & 0.183 & - & 0.227 & 0.214 &
0.974 & 1.677\\
$^{62}$Ga & 1.876 & 2.33  & 0.261 & 0.169 & - & 0.471 & 0.379 &
0.885 & 1.217\\
$^{66}$As & 0.848 & - & 0.066 & - & - & 0.499 & - &
0.911 & 1.236 \\
$^{70}$Br & 1.083 & - & 0.089 & - & - & 0.313 & - &
0.801 &  1.377 \\
$^{74}$Rb & 2.258 & - & 0.069 & - & - & 0.223 & - &
0.831 & 1.716 \\
\end{tabular}
\end{table}
\noindent $^a$ From Ref.~\cite{Hag94}

\noindent $^b$ From Ref.~\cite{Tow77}

\newpage
\begin{table}
\caption{Same as Table II for $A\ge 58$ using the FPVH interaction.}
\begin{tabular}{cccccccccc}
A & $E_{x,th}^1$ & $E_{x,exp}^1$ & $\delta_{IM}^1$ & $\delta_{IM,s}^1$ &
 $\delta_{IM}$ & $\delta_{IM,s}$ &
$\delta_{RO}^{{\rm HF}}$ &  $\delta_{RO}^{{\rm WS}}$\\
\tableline
$^{58}$Zn & 2.850 & 2.943 & 0.224 & 0.258 & 0.231 & 0.265 &
0.997 & 1.762 \\
$^{62}$Ga & 1.460 & 2.33  & 0.201 & 0.079 & 0.408 & 0.286 &
1.029 & 1.409 \\
$^{66}$As & 1.250 & - & 0.019 & - & 0.388 & - &
1.243 & 1.577 \\
$^{70}$Br & 1.545 & - & 0.017 & - & 0.330 & - &
1.082 & 1.596 \\
$^{74}$Rb & 2.988 & - & 0.090 & - & 0.237 & - &
0.670 & 1.409 \\
\end{tabular}
\end{table}

\newpage
\begin{table}
\caption{Comparion between theoretical and experimental excitation energies
(in MeV) of the first $J^\pi=2^+$ state in even-even $N=Z$ nuclei.}
\begin{tabular}{cccc}
A & FPVH & FPD6$^*$ & Exp \\
\tableline
$^{60}$Zn & 1.134 & 0.825 & 1.004$^a$ \\
$^{64}$Ge & 0.914 & 0.700 &  0.902$^c$ \\
$^{68}$Se & 0.939 & 0.600 &  0.854$^c$ \\
$^{72}$Kr & 0.976 & 0.707 &  0.709$^c$ \\
$^{76}$Sr & 0.892 & 0.752 &  0.261$^c$ \\
$^{80}$Zr & - & - &  0.289$^c$ \\
\end{tabular}
\end{table}
\noindent $^a$ from Ref.~\cite{Led78}.

\noindent $^c$ from Ref.~\cite{Lis88}.

\newpage
\begin{table}
\caption{List of isospin-mixing corrections $\delta_{IM}$,
$\delta_{RO}$, and $\delta_C$ (in \%), $f_R t$ and ${\cal F}t$ (in seconds)
for the accurately measured cases.}
\begin{tabular}{cccccc}
A & $\delta_{IM}$ & $\delta_{RO}^{{\rm HF}}$ &
$\delta_C$ & $f_Rt^c$ & ${\cal F}t$ \\
\tableline
$^{10}$C & 0.04 & 0.11 & 0.15(9) & 3154.4$\pm 5.1\pm 2.4$ & 3148.5(64) \\
$^{14}$O & 0.01$^a$ & 0.14 & 0.15(9) & 3151.1$\pm 1.8\pm 2.4$ & 3144.0(51) \\
$^{26m}$Al & 0.01$^a$ & 0.29 & 0.30(9) & 3157.8$\pm 1.7\pm 2.4$ & 3147.2(45) \\
$^{34}$Cl & 0.06$^a$ & 0.51 & 0.57(9) & 3167.0$\pm 1.9\pm 2.4$ & 3148.8(45) \\
$^{38m}$K & 0.11$^a$ & 0.48 & 0.59(9) & 3166.5$\pm 2.6\pm 2.4$ & 3146.3(49) \\
$^{42}$Sc & 0.11$^a$ & 0.31 & 0.42(9) & 3168.1$\pm 1.4\pm 2.4$ & 3148.7(46) \\
$^{46}$V & 0.09 & 0.29 & 0.38(9) & 3165.5$\pm 1.8\pm 2.4$ & 3151.6(46) \\
$^{50}$Mn & 0.02 & 0.33 & 0.35(9) & 3164.2$\pm 1.6\pm 2.4$ & 3149.6(56) \\
$^{54}$Co & 0.04 & 0.40 & 0.44(9)$^b$ & 3166.4$\pm 1.1\pm 2.4$ & 3152.8(46) \\
\end{tabular}
\end{table}
\noindent $^a$ from Ref.~\cite{Orm89}.

\noindent $^b$ using $\delta_{IM}=0.04(1)$ as discussed in the text.

\noindent $^c$ From the new Chalk River data set~\cite{Tow95}. The systematic
uncertainty
of 2.4~s is due to the systematic uncertainty of 0.08\% in
$\Delta_R$~\cite{Bar92}.

\end{document}